\documentstyle[prl,aps,epsf,floats]{revtex} \def\narrowtext{} \tighten
\twocolumn
\input epsf.sty

\begin{document}
\draft

\title{BCS-like Bogoliubov Quasiparticles in High-{\it T}$_c$ 
Superconductors Observed by Angle-Resolved Photoemission Spectroscopy
}

\author{
		H. Matsui$^1$,
		T. Sato$^1$,
		T. Takahashi$^{1}$,
		S.-C. Wang$^2$,
		H.-B. Yang$^2$,
		H. Ding$^2$,
		T. Fujii$^{3,*}$,
		T. Watanabe$^{4,\dagger}$,\\
		and A. Matsuda$^4$
        }
\address{
		(1)Department of Physics, Tohoku University, Sendai 
980-8578, Japan\\
		(2)Department of Physics, Boston College, Chestnut Hill, MA 
02467, USA\\
		(3)Department of Applied Physics, Faculty of Science, 
Science University of Tokyo, Tokyo 162-8601, Japan\\
		(4)NTT Basic Research Laboratories, Atsugi 243-0198, Japan
          }
\address{%
\begin{minipage}[t]{6.0in}
\begin{abstract}
\typeout{polish abstract}
  We performed high-resolution angle-resolved photoemission spectroscopy on 
triple-layered high-{\it T}$_c$ cuprate 
Bi$_2$Sr$_2$Ca$_2$Cu$_3$O$_{10+\delta}$.  We have observed the full energy 
dispersion (electron and hole branches) of Bogoliubov quasiparticles and 
determined the coherence factors above and below {\it E}$_F$ as a function of 
momentum from the spectral intensity as well as from the energy 
dispersion based on BCS theory.  The good quantitative agreement between 
the experiment and the theoretical prediction suggests the basic validity 
of BCS formalism in describing the superconducting state of cuprates.
\end{abstract}
\end{minipage}}
\maketitle
\narrowtext
It is well known that BCS (Bardeen, Cooper, and Schrieffer) theory 
\cite{BCS} explains many of fundamental thermodynamic, transport and 
magnetic properties of superconductors by introducing a simple picture that 
two electrons with opposite momenta and spins near the Fermi surface form a 
pair (Cooper pair) in the superconducting state.  In the language of Fermi 
liquids \cite{Landau}, the quasiparticles (QPs) of the Cooper pairs are 
called Bogoliubov quasiparticles (BQPs) \cite{Bogoliubov} which are defined 
as excitation of a single electron ``dressed" with an attractive 
interaction between paired electrons.  The BQPs play an essential role in 
characterizing the superconducting state via quantities such as the superconducting gap 
and its symmetry.  Despite the excellent description of the BQPs in BCS 
theory, there has been no direct experimental observation of the predicted 
full energy dispersion (electron and hole branches) of BQPs, although the 
pairing of electrons is evidenced by ac Josephson-junction experiment 
\cite{Jaklevic}.  For high-{\it T}$_c$ cuprates, the direct observation of 
the BQPs is even more significant, since many microscopic ideas of BCS 
theory have been seriously challenged.  Since the BQPs are a natural 
consequence of the starting Hamiltonian including the two-body attractive 
interaction assumed by BCS theory \cite{BCS}, the existence of BQPs would 
prove the validity of the basic BCS description of the superconducting 
state in the cuprates.

      In this Letter, we report the direct observation of BQPs in 
triple-layered high-{\it T}$_c$ cuprate 
Bi$_2$Sr$_2$Ca$_2$Cu$_3$O$_{10+\delta}$ by angle-resolved photoemission 
spectroscopy (ARPES).  By using ultrahigh resolution in energy and 
momentum, we have succeeded in directly observing the full energy 
dispersion (electron and hole branches) of BQPs below and above the Fermi 
level ({\it E}$_F$).  We have determined experimentally the coherence 
factors as a function of momentum from ARPES intensity and compared the 
result with the prediction from BCS theory to investigate the basic 
validity of the theory in high-{\it T}$_c$ cuprates.

\begin{figure}[!t]
\epsfxsize=3.4in
\epsfbox{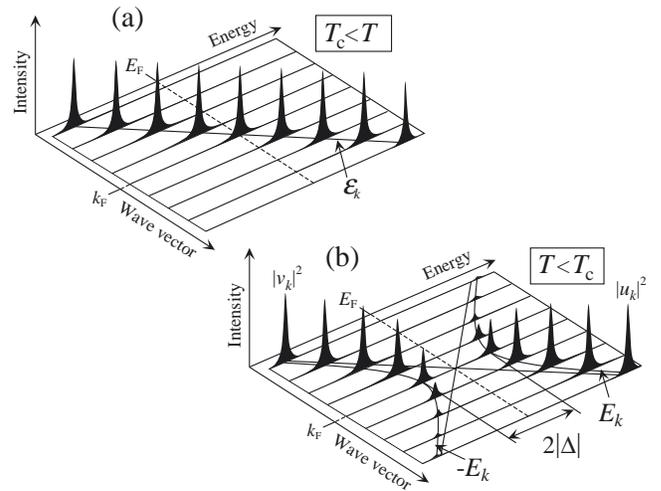}
\vspace{0.0cm}
\caption{Schematic diagram of the formation process of a Bogoliubov-quasiparticle 
(BQP) band.  In the normal state above ${\it T}_c$ (a), the electron band 
has an equal weight at any momentum, while in the superconducting state 
below ${\it T}_c$ (b), particle-hole mixing (mixing of electron and hole 
bands) takes place due to the pairing, leading to opening of a 
superconducting gap as well as a transfer of weight between the electron 
and hole bands.}
\label{Fig. 1}
\end{figure}

\begin{figure*}[!t]
\epsfxsize=7.1in
\epsfbox{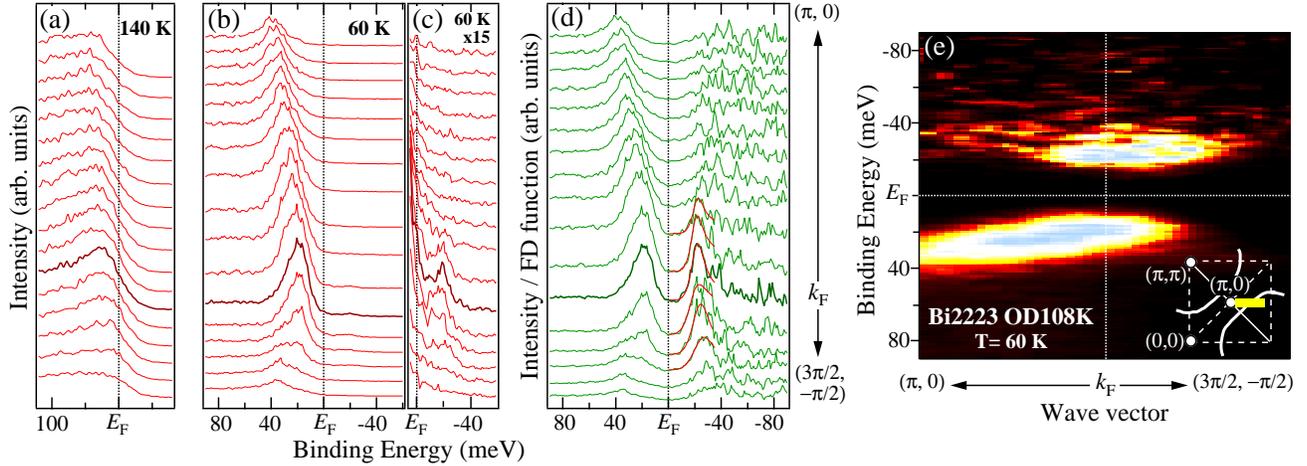}
\vspace{0.0cm}
\caption{(a) ARPES spectra of Bi2223 in the normal state (140 K) 
measured along a yellow line in the Brillouin zone shown in the inset in 
(e).  Spectrum at ${\it k}_F$ is indicated by a dark red line.  
(b) ARPES spectra taken under the same condition as in (a) but at the superconducting state (60 K).
(c) Same as (b) in an expanded intensity scale above ${\it E}_F$.
(d)  ARPES spectra in (b) divided by the FD function at 60 K convoluted with a 
Gaussian reflecting the instrumental resolution. Spectrum at
 ${\it k}_F$ is indicated by a dark green line.  Red lines are fitting
 curves for unoccupied states using a Lorentzian with energy-dependent broadening
 factor.  Fittings are restricted to the spectra of which peak positions are located
 within 5$k_B${\it T} from ${\it E}_F$.
  (e) Intensity plot of normalized ARPES spectra in (d) as a function of 
binding energy and wave vector.  Momentum region is the same in (a)-(d).
}
\label{Fig. 2}
\end{figure*}

      High-quality single crystals of 
Bi$_2$Sr$_2$Ca$_2$Cu$_3$O$_{10+\delta}$ (Bi2223, overdoped, {\it T}$_c$=108 K) were 
grown by the traveling solvent floating zone method \cite{Fujii}. 
ARPES measurements were performed using a GAMMADATA-SCIENTA SES-200 
spectrometer with 22 eV photons at the undulator 4m-NIM beamline at 
Synchrotron Radiation Center in Wisconsin.  The energy and angular 
(momentum) resolutions were set at 11 meV and 0.2$^{\circ}$ 
(0.007\AA$^{-1}$), respectively.  A clean fresh surface of sample was 
obtained by {\it in-situ} cleaving along the (001) plane.  The Fermi level 
({\it E}$_F$) of the sample was referenced to that of a gold film evaporated 
onto the sample substrate.

      Figure 1 shows a schematic comparison of spectra between the 
superconducting BQPs and the normal-state QPs.  In the normal state [Fig. 
1(a)], one-electron like QPs disperse across {\it E}$_F$ with an equal 
spectral weight, defining the Fermi vector ({\it k}$_F$) at {\it E}$_F$. 
In the superconducting state [Fig. 1(b)], the BQPs split the single band 
into two dispersing branches below and above {\it E}$_F$.  The BQPs exhibit 
several characteristic features: (i) the dispersion of two branches is 
symmetric with respect to {\it E}$_F$; (ii) the minimum separation at {\it 
k}$_F$ defines the superconducting energy gap (2$\Delta$); (iii) both the 
dispersions below and above {\it E}$_F$ show bending-back behavior at {\it 
k}$_F$ due to the particle-hole mixing; (iv) the spectral weight of the two 
branches differs, and changes with the wave vector {\it k}; however, (v) 
the combining weight of the two branches is always constant at any ${\it 
k}$.

      Figures 2(a) and 2(b) show the ARPES spectra of Bi2223 in the normal and
superconducting states, respectively, measured along the 
direction in the Brillouin zone shown in the inset in Fig. 2(e). We have 
surveyed many areas in the Brillouin zone and confirmed that the intensity 
of Umklapp bands is very weak and they do not contaminate the main-band 
dispersion in this momentum region. While we see a broad peak dispersing across
{\it E}$_F$ in the normal state, we clearly observe a sharp coherent 
peak in the superconducting state\cite{Multi}.
As {\it k} is changed from ($\pi$, 0) to (3$\pi$/2, -$\pi$/2), the coherent 
peak gradually disperses toward {\it E}$_F$, showing a minimum energy gap 
at {\it k}$_F$ [dark red line in Fig. 2(b)], and then disperses back to the 
higher binding energy while rapidly reducing its intensity.  These 
behaviors are consistent with the band dispersion below {\it E}$_F$ shown 
in Fig. 1(b) as well as with a previous ARPES result on 
Bi$_2$Sr$_2$CaCu$_2$O$_{8+\delta}$ (Bi2212) \cite{Campuzano}.  More 
importantly, we find additional weak but discernible structures about 20 
meV above {\it E}$_F$ in the spectra, which are more clearly seen in Fig. 
2(c).  This new structure shows a clear momentum dependence with a stronger 
intensity in the region of $|k| > | k_F|$, opposite to the behavior of the 
band below {\it E}$_F$.  We ascribe these small structures 
to the BQP band above {\it E}$_F$ by referring to Fig. 1(b) \cite{condition}.  The 
considerably weak intensity of the peaks above {\it E}$_F$ is simply due to 
the effect of the Fermi-Dirac (FD) function.  In 
order to see more clearly the band dispersion above {\it E}$_F$, we have 
divided the ARPES spectra by the FD function at 60 K convoluted with the 
instrumental resolution \cite{Greber}.  The result is shown in Fig. 2(d), 
where we find a dispersive structure with comparable intensity above {\it 
E}$_F$, although the signal-to-noise ratio is relatively low because of the 
originally small ARPES intensity.  It is remarked here that dispite
 the low signal-to-noise ratio the bending-back behavior of band is also seen
 in the unoccupied states as in the occupied states.  The dispersive
 feature is more clearly visualized in Fig. 2(e) by plotting the renormalized ARPES
 intensity as a function of momentum and energy.  We observe several 
characteristic behaviors for the two branches: (i) the dispersive feature 
is almost symmetric with respect to {\it E}$_F$ while the intensity is not; 
(ii) the bands have a minimum energy gap at {\it k}$_F$; (iii) both bands 
show the bending-back effect at {\it k}$_F$; (iv) the spectral intensity of 
the two bands show opposite evolutions as a function of {\it k} in the 
vicinity of {\it k}$_F$. All these features qualitatively agree with the 
behaviors of BQPs predicted from BCS theory [Fig. 1(b)], suggesting the 
basic validity of the BQP concept in high-{\it T}$_c$ superconductors.

      The next question is how quantitatively the observed dispersion and 
spectral weight agree or disagree with the predictions from BCS theory.  This 
point is crucial in establishing the BQP concept more firmly in the 
cuprates.  Figure 3(a) shows direct comparison between the experimental 
band dispersion of BQPs (same as Fig. 2(e)) and the corresponding 
calculated curves based on BCS theory.   In BCS theory, the band dispersion 
of BQPs ($E_k$) is expressed as $E_k 
=[{\epsilon_k}^2+|\Delta(k)|^2]^{1/2}$, where $\epsilon_k$ and $\Delta(k)$ 
are the normal-state dispersion and the superconducting gap, respectively. 
We have determined at first $\epsilon_k$ and $\Delta(k)$ from the ARPES 
spectra at 140 and 60 K in Fig. 2, respectively, and then calculated the 
``theoretical" band dispersion of BQPs using the above equation.  The 
normal state dispersion $\epsilon_k$ [white solid line in Fig. 3(a)] is 
obtained from the peak position in ARPES spectra at 140 K [white open circles in 
Fig. 3(a)] by fitting them with parabolic function.  The superconducting 
gap  $\Delta(k)$ is assumed to be the $d_{x^2-y^2}$-wave superconducting 
order parameter $\Delta(k)=\Delta_0|cos(k_x)-cos(k_y)|/2$, where $\Delta_0$ 
is determined by the peak energy of the 60-K spectrum at {\it k}$_F$. We 
find in Fig. 3(a) that the calculated dispersion well traces the 
strong-intensity areas of ARPES spectra, showing good agreement in the 
band dispersion between the experiment and the theory. 
To further confirm the agreement in the unoccupied states, we have fit the
 normalized ARPES spectra in Fig. 2(d) with a Lorentzian and show the peak
 positions in Fig. 3(a).  The peak positions are found to be well on the calculated
 curve, showing quantitatively good agreement between the experiment
 and the calculation.

\begin{figure}[!t]
\epsfxsize=3.4in
\epsfbox{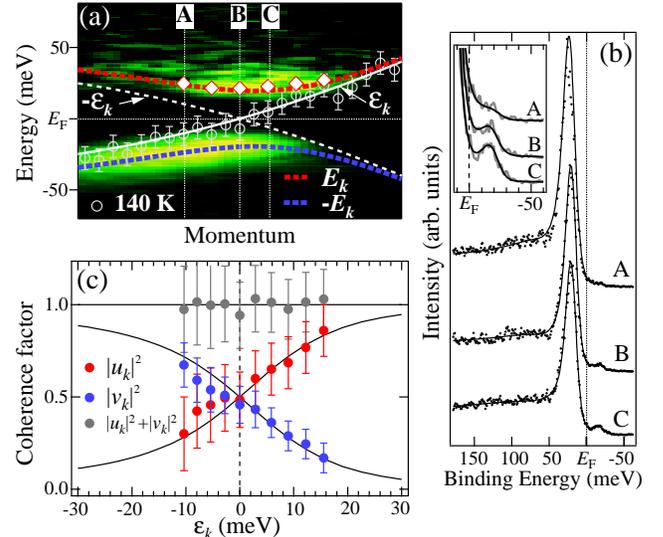}
\vspace{-0.2cm}
\caption{(a) Comparison of the ARPES intensity plot (same as Fig. 2(e)) 
with the calculated BQP band dispersions based on BCS theory (red and blue 
dashed lines).  Calculation has been performed using the BCS formula $E_k 
=[{\epsilon_k}^2+|\Delta(k)|^2]^{1/2}$, where $\epsilon_k$ (white solid 
line) has been determined by fitting the spectral peak positions at 140 K 
(white open circles) using parabolic function. The corresponding hole band 
($-\epsilon_k$) is also shown by a white dashed line.  Peak
 positions of Lorentzian to fit the normalized ARPES spectra in the unoccupied states (Fig. 2(d))
 are shown by white filled squares for comparison.
(b) ARPES spectra (dots) measured at three representative points in the 
Brillouin zone shown in (a) (points A, B, and C), together with the fitting 
curves (solid lines) to obtain the coherence factors ($|u_k|^2$ and 
$|v_k|^2$).  The inset shows expansion near $E_F$.
(c) Coherence factors as a function of $\epsilon_k$ determined from ARPES 
intensity (red and blue solid circles), compared with those derived from 
ARPES dispersion using BCS formula (solid lines).  Sum of two 
coherence factors ($|u_k|^2$+$|v_k|^2$) obtained by ARPES intensity is 
shown by gray solid circles.
}
\label{Fig. 3}
\end{figure}

      To further study the validity of the BQP concept, we have determined 
the coherence factors above and below {\it E}$_F$, $|{\it u}_k|^2$ and $|{\it 
v}_k|^2$, by using the following two independent methods and compared them 
with each other.  The coherence factors are defined as the relative 
intensity of BQP bands above and below {\it E}$_F$, which directly correspond 
to the normalized ARPES spectral intensity in the superconducting state. 
On the other hand, according to BCS theory, the coherence factors can be 
also calculated from the energy dispersions of the normal-state QP and the 
superconducting BQP bands in the following way:
	 \begin{equation}
	 |{\it u}_k|^2=1-|{\it v}_k|^2=\frac{1}{2}\:(1+\frac{\epsilon_k}{{\it E}_k})
	 \end{equation}
where $\epsilon_k$ and $E_k$ are the energy of the normal QP and BQP bands, 
respectively \cite{BCS}.

      For the first method, we have determined $|{\it u}_k|^2$ and $|{\it 
v}_k|^2$ by fitting the original ARPES spectra in Fig. 2(b) by the 
following equation,
	 \begin{eqnarray}
&&I(k,\omega)=\nonumber\\
&&I_0(k){\int\!\!d\omega'\{A_{BCS}(k,\omega')\!+\!A_{inc}(k,\omega')\}
f(\omega')R(\omega\!-\!\omega')}
	\end{eqnarray}
where $I_0(k)$ is a prefactor which includes some kinematical factors and 
the dipole matrix-element.  $A_{BCS}(k, \omega)$ is the BCS spectral 
function expressed as,
	\begin{equation}
A_{BCS}(k, \omega) = \frac{1}{\pi}\{\frac{|u_k|^2\Gamma}{(\omega-E_k)^2+ 
\Gamma^2}+\frac{|v_k|^2 \Gamma}{(\omega+E_k)^2+ \Gamma^2}\}
	\end{equation}
where $\Gamma$ is a linewidth broadening due to the finite lifetime of 
photoholes.  $A_{inc}(k, \omega)$ in Eq. (2) is an empirical function 
representing the incoherent background \cite{Bkw},  $f(\omega)$ is the 
FD function, and $R(\omega)$ is the Gaussian
resolution function.  To remove the effect of $I_0(k)$, we have 
divided the spectral intensity of the superconducting state (60 K) with the 
integrated normal-state (140 K) spectral intensity at each {\it k} point. 
We determine the peak weights below and above {\it E}$_F$ at each {\it k} 
point by decomposing the spectrum, and then divide them 
by the average value of the total peak weight at each {\it k} point 
\cite{Ideally}. We define these normalized weights as $|{\it v}_k|^2$ and 
$|{\it u}_k|^2$ and show the results in Fig. 3(c).  It is important to note 
here that $|{\it u}_k|^2$ and $|{\it v}_k|^2$ are determined independently 
without using the sum rule of $|{\it u}_k|^2$+$|{\it v}_k|^2$=1.  The 
fittings have been performed within the momentum region where 
reliable values for the coherence factors can be obtained [shown in Fig. 
3(c)], since the reliability of fitting is considerably decreased 
far above {\it E}$_F$ due to the low signal-to-noise ratio of the original 
ARPES spectra.

      For the second method, we have calculated $|{\it u}_k|^2$ and $|{\it 
v}_k|^2$ by using Eq. (1) with $\epsilon_k$ and $E_k$ obtained from 
Fig. 3(a).  In Fig. 3(c), we show the comparison of the two sets of 
coherence factors determined by the ARPES intensity itself and by the BCS 
calculation based on the band dispersions, respectively. We find a 
surprisingly good quantitative agreement between the two sets of coherence 
factors determined totally independently \cite{Aslight}.  It is also 
remarked that the sum of coherence factors, $|{\it u}_k|^2$+$|{\it 
v}_k|^2$, determined by the ARPES intensity is almost constant over the 
measured momentum region, in good agreement with the sum rule of the 
coherence factors predicted from BCS theory.

      The sharp superconducting quasiparticle peak at ($\pi$, 0) in Bi-based 
cuprates has been widely studied in the high-{\it T}$_c$ field. The origin 
and its implications are still under debate. Some of its properties 
clearly deviate from BCS theory.  For example, the leading edge of the 
quasiparticle peak does not reach {\it E}$_F$ even above {\it T}$_c$, 
indicating the opening of a pseudogap \cite{DingPG}.  Spectral weight of 
the quasiparticle peak increases with hole-doping and has an unusual 
temperature dependence \cite{Feng,Ding}.  The present ARPES study, on the 
other hand, by its direct experimental observation of two dispersive 
branches with their dispersions and coherence factors consistent with the 
BCS predictions, has unambiguously established the Bogoliubov-quasiparticle 
nature of this sharp peak. This implies that the superconducting state of 
the high-{\it T}$_c$ cuprate is BCS-like, so that some of the basic BCS 
formalism is still valid in this case, although the pairing mechanism and 
other exotic properties are beyond BCS theory. Thus this experimental 
observation provides a strong constraint in modeling the high-{\it T}$_c$ 
problem. Moreover, this observation opens a new way to study quasiparticles 
in the superconducting state by probing the branch above {\it E}$_F$ 
populated by thermal or optical excitations.  One particularly important 
application is to determine the nature of the gapped peak in the pseudogap 
state. This may potentially settle the debate between the precursor pairing 
gap and single-particle gap.

We thank T. Tohyama, K. Yamada, J. R. Engelbrecht and Z. Wang for useful 
discussions.  This work is supported by grants from the MEXT of Japan, US 
NSF DMR-0072205, and Sloan Foundation. T.S. thanks JSPS for financial support.  
The Synchrotron Radiation Center is supported by US NSF DMR-0084402.

$^*$Present address: Department of Applied Physics, Waseda University, 
Tokyo 169-8555, Japan \\
$^{\dagger}$Present address: NTT Photonics Laboratories, Atsugi 243-0198, Japan

\newpage


\begin{references}
\bibitem{Multi}In the present study of Bi2223, multiple bands are not 
clearly resolved either in the normal or in the superconducting state [7,8] in 
contrast to the case of Bi2212.
\bibitem{Ideally}  We did not use the normal-state quasiparticle weight for 
division, because the quaspiarticle weight in ARPES spectrum shows 
temperature dependence in cuprates [13,14].
\end{references}
\end{document}